\documentclass[twocolumn,showpacs]{revtex4}
\usepackage[dvips]{graphicx}
\usepackage{dcolumn}
\usepackage{amsmath}
\makeatletter
\def\btt#1{\texttt{\@backslashchar#1}}%
\DeclareRobustCommand\bblash{\btt{\@backslashchar}}%
\makeatother
\begin{document}

\title{Dynamics of a dislocation bypassing an impenetrable precipitate: the Hirsch mechanism revisited}
\author{Takahiro Hatano}
\affiliation{
Earthquake Research Institute, University of Tokyo, Tokyo 113-0032, Japan.}
\date{\today}

\begin{abstract}
Dynamical process where an edge dislocation in fcc copper bypasses an impenetrable precipitate 
is investigated by means of molecular dynamics simulation.
A mechanism which is quite different from the Orowan mechanism is observed, 
where a dislocation leaves two prismatic loops near a precipitate: i.e. the Hirsch mechanism.
It is found that the critical stress for the Hirsch mechanism is almost the same as the Orowan stress, 
while the spatial inhomogeneity of the shear stress is essential to the Hirsch mechanism.
We also find that the repetition of the Hirsch mechanism does not increase the critical stress.
\end{abstract}

\pacs{81.40.Cd, 61.72.Bb, 62.20.Fe}
\maketitle 

Plastic flow in crystalline materials is mainly dominated by the movement of dislocations.
They are curvilinear defects which can move at much lower stress levels than 
the theoretical strength of a perfect crystal \cite{friedel}.
A dislocation interacts with other lattice defects such as voids or precipitates, 
which obstruct the dislocation motion to result in hardening.
Such dislocation-obstacle interactions dominate plasticity of crystalline materials, 
and hence they are one of the major concerns in materials science.
Until very recently, they were considered exclusively by plausible continuum models 
\cite{bacon,scattergood}, which deal with a single glide plane and neglect atomistic discreteness.
The recent computer development enables direct molecular dynamics (MD) simulation 
on dislocation systems, which consist of multi-million atoms.
MD simulations have revealed dynamical properties and atomistic details of dislocations.
For example, edge dislocations absorb vacancies when they interact with voids \cite{osetsky} 
or stacking fault tetrahedra \cite{wirth}.
They are good illustrations of the usefulness of MD simulation in the field of dislocation physics.
Along the line of these studies, the interaction with an impenetrable precipitate, 
which has not been investigated by MD simulation, is explored in this Letter.

An interaction between a dislocation and a precipitate is characterized by their shear moduli.
When the shear modulus of a precipitate is larger than the bulk's, the interaction is repulsive 
so that the precipitate can be impenetrable \cite{incoherency}.
In this case, the extent of hardening becomes significant.
This is the principle of so-called "particle strengthening", which is widely utilized in processing 
stronger materials \cite{argon}.
There is a mechanism by which a dislocation bypasses impenetrable obstacles, 
where a dislocation largely bows out to leave a dislocation loop around the obstacle \cite{friedel}.
Note that the loop is on the original glide plane.
This process and the resultant dislocation loop are referred to as the Orowan mechanism 
and an Orowan loop, respectively.
However, because the interaction between an Orowan loop and a dislocation is strongly repulsive, 
theoretical calculations in which Orowan loops are assumed to be persistent 
predict unreasonably strong work-hardening \cite{ashby,brown1}.
To avoid this contradiction, some alternative bypass mechanisms have been presented.
Hirsch \cite{hirsch1,humphreys} had pointed out the possibility of prismatic loops formation in a bypass mechanism.
This mechanism is named after Hirsch and referred to as the Hirsch mechanism.
On the other hand, Brown and Stobbs presented another mechanism where a dislocation loop 
of a secondary Burgers vector nucleates on the precipitate surface \cite{brown2}.
In both cases, the resultant structure is a row of prismatic loops which was experimentally observed 
by transmission electron microscopy (TEM) \cite{humphreys,brown2,sarnek}.
However, since the dynamical process cannot be observed by TEM,
the postulated mechanisms have not been directly tested.
In this Letter, we wish to clarify the dynamics of prismatic loops formation 
and to find out quantitative conditions which determine resultant dislocation loops: 
Orowan loops or prismatic loops. Along the line of thought, an MD simulation is performed.


Let us describe the computational model.
We consider fcc copper utilizing a many-body interatomic potential of Finnis-Sinclair \cite{finnis} 
and adopt parameters which are determined by Ackland et al \cite{ackland}.
The lattice constant $a=3.615$ \AA.
The dimensions of the model system are $x=23\times[11\bar{2}]$ ($20$ nm), 
$y=79\times[1\bar{1}0]$ ($40$ nm), and $z=27\times[111]$ ($17$ nm).
This system consists of approximately $1.2$ million atoms.
(In addition to this system, we prepare some smaller systems which have different length in the $x$ direction.)
We introduce an edge dislocation parallel to the $x$ axis, setting the Burgers vector to be parallel to the $y$ axis.
Note that we focus on an edge dislocation, since a screw dislocation can avoid a precipitate 
by switching its glide planes via cross-slip \cite{note1}, which is not our interest here.
An impenetrable precipitate is modeled by a set of immobile atoms, which are coherent to the bulk crystal.
It mimics a precipitate of the infinite shear modulus, which cannot be sheared by dislocations.
Periodic boundary conditions are employed in the $x$ and the $y$ directions,
and the surfaces exist only in the $\pm z$ directions.
In order to cause the shear strain, $(111)$ surface is displaced at a constant velocity.
We test two cases, which are referred to as conditions I and II, respectively.
In condition I, the upper ($+z$) surface move towards the $-y$ direction 
while the lower surface is fixed.
In condition II, both surfaces move opposite to each other.
In both conditions, we set the strain rate $\dot{\epsilon}=7 \times 10^6$ [$\rm{s}^{-1}$].
Note that we adopt the condition I unless explicitly indicated.
Temperature is set to be $300$ K, where the velocities of copper atoms are given randomly 
according to the Maxwell-Boltzmann distribution.

\begin{figure}
\includegraphics[scale=0.7]{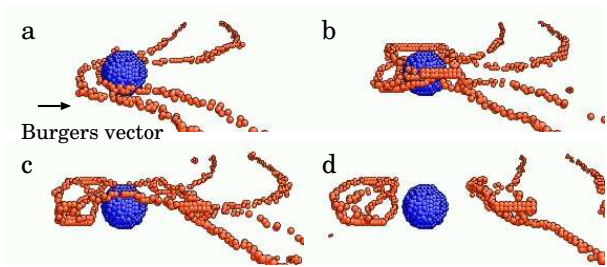}
\caption{Successive snapshots of an edge dislocation bypassing an impenetrable precipitate.
The radius of the precipitate (blue atoms) is $1.5$ nm.
A dislocation (red atoms) is visualized by omitting atoms which have $12$ nearest neighbors.
(a) A dislocation bows out to form a screw dipole. 
(b) The screw dipole cross-slips.
(c) The screw dipole undergoes double cross-slip to annihilate each other at the top of the precipitate.
(d) The dislocation has bypassed the precipitate with a superjog. A prismatic loop is left behind.}
\label{snapshots}
\end{figure}

\begin{figure}
\includegraphics[scale=0.5]{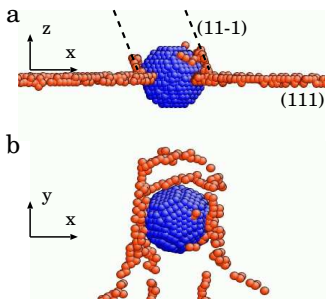}
\caption{Geometry of the primary cross-slip from $(111)$ to $(11\bar{1})$.
(a) A $[1\bar{1}1]$ projection. (b) A $[111]$ projection.}
\label{crossslip}
\end{figure}

\begin{figure}
\begin{center}
\includegraphics[scale=0.43]{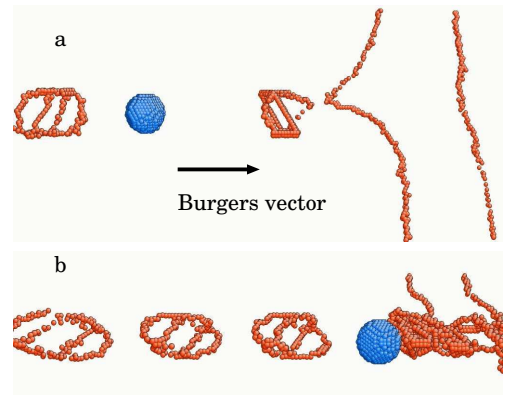}
\caption{(a) Configuration just after a single bypass process.
A dislocation leaves two prismatic loops: an interstitial loop and a vacancy loop.
The secondary loop is formed relatively far from the precipitate, 
since it has been dragged by the dislocation as a superjog.
(b) A row of prismatic loops after the passage of three dislocations.
Note that the loops on the right side of the precipitate do not form an apparent row.}
\label{loops}
\end{center}
\end{figure}

Let us turn to the simulation results. First, we focus on the dynamics and the geometry.
The configuration is shown in Fig. \ref{snapshots} \cite{li}, 
where a dislocation does not leave an Orowan loop but two prismatic loops.
At the first stage, a dislocation bows out to form a screw dipole of opposite signs.
The main difference between the Orowan mechanism and the mechanism observed here 
lies in the behavior of this screw dipole.
It cross-slips onto a $(11\bar{1})$ plane (see FIG. \ref{crossslip}), 
and then undergoes double cross-slip onto another $(111)$ plane,
on which the screw dipole eventually annihilates each other.
In FIG. \ref{snapshots} (d), a prismatic loop is formed on the left side of the precipitate 
and a superjog is formed on the opposite side.
Then the superjog is dragged by the dislocation for a while, and is eventually separated to 
form the secondary prismatic loop as shown in Fig. \ref{loops} (a).
If the primary cross-slip takes place towards the extra atomistic half plane 
which constitutes the edge dislocation, the prismatic loop created before the precipitate is interstitial
and the secondary loop after the precipitate is vacancy.
(If the primary cross-slip goes opposite, the order is reversed).
Note that the resultant prismatic loops have the same Burgers vector as that of the original edge dislocation.
Indeed, this mechanism is identical to the Hirsch mechanism \cite{humphreys}.
We stress that this is the first case where the Hirsch mechanism is dynamically observed, 
while TEM observations have detected only the resultant structures.
Note that the mechanism of Brown and Stobbs \cite{brown2} has not been observed 
in the conditions investigated here.
More importantly, it is found that the Hirsch mechanism always occurs for the condition I, 
while the Orowan mechanism is realized for the condition II.
To clarify the underlying physics, we have to investigate the quantitative aspect.

\begin{figure}
\includegraphics[scale=0.5]{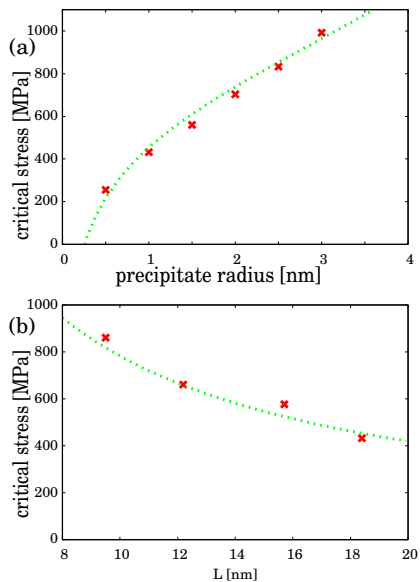}
\caption{The critical resolved shear stresses for the Hirsch mechanism (denoted by the red $\times$ symbols).
Note that the Orowan stress is almost the same as the Hirsch stress.
The green dashed lines represent Eq. (\ref{logarithmiclaw}) with $A=6.5$ N/m and $B=0.5$ nm.
(a) Dependence of the critical stress on the precipitate radius, where $L_x$ is $20.4$ nm.
(b) Dependence of the critical stress on the spacing of the precipitates, where $r$ is $1.5$ nm.}
\label{r-crss}
\end{figure}

We investigate the critical resolved shear stress for the Hirsch mechanism, which we will call the "Hirsch stress".
The shear stress is defined as the area-averaged force acting on the surfaces.
The Hirsch stress is defined as the maximum shear stress during a bypass process.
It is realized just before the primary cross-slip, by which the Hirsch stress is determined.
Behaviors of the Hirsch stress with respect to the precipitate radius and to the precipitate spacing
are shown in Fig. \ref{r-crss}, where we find the Hirsch stress is described by the following logarithmic law.
\begin{equation}
\label{logarithmiclaw}
\sigma_{yz}=\frac{A}{L}\log\frac{1}{\left(0.5r^{-1}+L^{-1}\right)B}, 
\end{equation}
where $A$ and $B$ denote material-dependent parameters, $r$ is the radius of a precipitate, 
and $L$ is the precipitate spacing in the $x$ direction.
Namely, $L=L_x-2r$, where $L_x$ denotes the system size in the $x$ dimension.
Note that $A=6.5$ N/m and $B=0.5$ nm in the present system.
This kind of logarithmic behavior of the critical stress is universal in the context of 
dislocation-obstacle interactions \cite{osetsky,hatano}, the reason of which is explained as follows.
From the balance of forces that are transverse to the dislocation line, 
the resolved shear stress ($\tau$) and the radius of curvature ($\rho$) of a bowing dislocation are interrelated as 
\begin{equation}
\label{balance}
\tau=\frac{\gamma}{\rho b}.
\end{equation}
Note that $\gamma$ denotes the line tension, which depends on the configuration of a bending dislocation.
When an edge dislocation bends to form a screw dipole at a pinning point, the line tension is estimated as \cite{bacon} 
\begin{equation}
\label{gamma}
\gamma=\frac{Gb^2}{4\pi}\log \frac{1}{B}\left(\frac{1}{2r}+\frac{1}{L}\right)^{-1}.
\end{equation}
Since a fully bending dislocation forms a semi-arc between two pinning points, the radius of curvature is $\rho=L/2$.
Inserting this relation and Eq. (\ref{gamma}) into Eq. (\ref{balance}) immediately leads to the logarithmic law.
\begin{equation}
\label{logarithmiclaw2}
\tau=\frac{Gb}{2\pi L}\log\frac{1}{B}\left(\frac{1}{2r}+\frac{1}{L}\right)^{-1}.
\end{equation}
Although Eqs. (\ref{logarithmiclaw}) and (\ref{logarithmiclaw2}) are of the same form, 
the quantitative discrepancy is considerable.
Namely, $Gb/2\pi=1.96$ N/m in Eq. (\ref{logarithmiclaw2}), while the corresponding factor 
in Eq. (\ref{logarithmiclaw}) is $A=6.5$ N/m, which is three times larger than $Gb/2\pi$.
There are at least two reasons for this disagreement.
The first is that dissociation of dislocations is not taken into account in deriving Eq. (\ref{logarithmiclaw2}).
It takes much more stress to bend two partial dislocations simultaneously \cite{hatano}.
The second is the high dislocation density.
Since the glide of a single dislocation causes large strain relaxation in small systems, 
the critical depinning stress may depend on the initial position of a dislocation in MD simulations.

Then we discuss the difference between the Hirsch and the Orowan mechanisms, 
both of which are realized in the present simulation depending on the boundary conditions.
It should be remarked that the Orowan stress is almost the same as the Hirsch stress, 
and therefore we must refine the analysis.
Recalling that the Hirsch mechanism is realized by displacing only the upper surface, 
inhomogeneity of the shear strain may play a key role.
In order to see the inhomogeneity more explicitly, we calculate the average atomistic displacements 
in the $y$ direction for each mechanism. We define 
\begin{equation}
\delta y(z)=\frac{1}{n(z)}\sum_i\left[y_i(t)-y_i(0)\right],
\label{deltay}
\end{equation}
where the subscript $i$ denotes the $i$-th atom.
We take the sum regarding all the atoms that belong to the layer of $z<z_i(t)<z+1.89$ \AA.
Then $n(z)$ is the number of atoms in the layer, and $t$ is the time just before each bypass mechanism begins 
(i.e. when the shear stress becomes maximum.)
\begin{figure}
\includegraphics[scale=0.3]{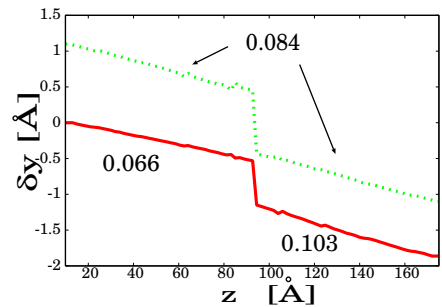}
\caption{Spatial profiles of the average atomistic displacement in the $y$ direction when the shear stress becomes maximum.
The red solid line represents $\delta y (z)$ for the Hirsch mechanism,
while the green dashed line denotes that for the Orowan mechanism. The gradient of $\delta y(z)$ for the Hirsch mechanism
is asymmetric with respect to $z=95$ \AA (the glide plane). System parameters $r=1.5$ nm and $L=20.4$ nm.}
\label{dyprofile}
\end{figure}
FIG. \ref{dyprofile} shows $\delta y(z)$ for each mechanism.
For the Hirsch mechanism, the gradient of $\delta y$ is different with respect to the glide plane.
Because the cross-slip which precedes the Hirsch mechanism occurs towards the larger stress region, 
it is concluded that the strong spatial inhomogeneity of the shear stress is essential to the Hirsch mechanism.
This spatial inhomogeneity is attributed to the existence of the precipitate.
Because the precipitate consists of immobile atoms, it inhibits the deformation caused by the boundary displacement.
Therefore, if the upper (or lower) surface is displaced, the elastic strain tends to localize on that side \cite{asymmetry}.
When the both surface is equivalently displaced, the elastic strain becomes symmetric to realize the Orowan mechanism.
The above discussion further leads us to the speculation that Orowan loops become unstable for the larger shear strain.
They can be decomposed into two prismatic loops via double cross-slip of the screw component, 
which is essentially the same manner as the Hirsch mechanism.
For example, in the system shown in FIG. \ref{dyprofile}, the shear strain of $0.103$ is enough to destabilize Orowan loops.

While the discussions so far involve a single bypass process, for the rest of this Letter,
iteration of the Hirsch mechanism is briefly investigated.
Namely, we consider the dynamics of subsequent dislocations which successively interact with a precipitate.
It involves the relevance of the Hirsch mechanism to work-hardening.
Although prismatic loops can move under an appropriate stress field (prismatic punching), 
they cannot move in the simple shear condition to be stagnant near the precipitate.
However, they do not cause hardening since subsequent dislocations can easily penetrate them.
For example, when $L_x=20.4$ nm and $r=1.5$ nm, it takes less than $180$ MPa for dislocations 
to penetrate prismatic loops, which is much smaller than the Hirsch stress itself, $560$ MPa.
Plastic flow is maintained by repetition of the Hirsch mechanism,
which yields a row of prismatic loops as is shown in Fig. \ref{loops} (b).
This is what has been observed in the experiments \cite{humphreys,brown2,sarnek}.
By contrast, the loops on the right side of the precipitate do not form an apparent row.
Because the prismatic loops move towards the $+y$ direction (the left in FIG. \ref{loops})
when they are sheared by dislocations, the loops on the $-y$ (the right) side of the precipitate
cannot move across the precipitate to collide each other there.
This phenomenon may be related to plastic cavitation which is often observed near precipitates \cite{sato}, 
and also to the rotational plastic flow around a nondeformable particle \cite{brown3}.

To conclude, the present study indicates that the dynamics of dislocation-obstacle interaction is much richer than considered before.
Cross-slip plays the crucial role in a bypass mechanism which is quite different to the Orowan mechanism.
We have clarified the condition that determines which bypass mechanism occurs.
Namely, the shear stress which is asymmetric with respect to the glide plane causes cross-slip of the screw dipole
and the Hirsch mechanism follows.
Further development along the line of the present study will be the relevance of the Hirsch mechanism 
to macroscopic plasticity. A three-dimensional continuum approach, which is referred to as "dislocation dynamics", 
is now developing to compute the many-body dynamics of dislocations and obstacles \cite{zbib}.
It may be friutful that the present results are suitably incorporated to such continuum approaches.
In addition, it should be remarked that the dislocation dynamics is closely related to the interface dynamics 
in the quenched disorder, which is extensively investigated in the field of nonlinear physics \cite{kardar}.
Dislocation physics would be much richer in communication with the different research field.

The author gratefully acknowledges helpful discussions with Yuhki Satoh, Hideki Matsui, and Hideo Kaburaki.


\begin{thebibliography}{99}
\bibitem{friedel}
J. Friedel, {\it Dislocations} (Pergamon Press, Oxford, 1964).

\bibitem{bacon}
D. J. Bacon, U. F. Kocks, and R. O. Scattergood, 
Phil. Mag. {\bf 28}, 1241 (1973).

\bibitem{scattergood}
R. O. Scattergood and D. J. Bacon, Acta Metall. {\bf 30}, 1665 (1982).

\bibitem{osetsky}
Yu. N. Osetsky and D. J. Bacon, J. Nucl. Mater. {\bf 323}, 268 (2003).

\bibitem{wirth}
B. D. Wirth, V. V. Bulatov, and T. Diaz de la Rubia,
J. Eng. Mater. Tech. {\bf 124}, 329 (2002).

\bibitem{incoherency}
Although incoherency is an important ingredient that causes impenetrability in many alloy systems, 
we do not consider incoherent precipitates here.

\bibitem{argon}
A. S. Argon, {\it Physics of Strength and Plasticity} (MIT Press, Cambridge, MA, 1969).

\bibitem{ashby}
M. F. Ashby, in {\it Oxide Dispersion Strengthening}, proceedings of the Second Bolton Landing Conference,
Bolton Landing, New York, 1966, edited by G. S. Ansell, T. D. Cooper, and F. V. Lenel
(Gordon and Breach, New York, 1968), p. 143.

\bibitem{brown1}
L. M. Brown and W. M. Stobbs, Phil. Mag. {\bf 23}, 1185 (1971).

\bibitem{hirsch1}
P. B. Hirsch, J. Inst. Metals {\bf 86}, 7 (1957).

\bibitem{humphreys}
F. J. Humphreys and P. B. Hirsch, Phil. Mag. {\bf 318}, 73 (1970).

\bibitem{brown2}
L. M. Brown and W. M. Stobbs, Phil. Mag. {\bf 23}, 1201 (1971).

\bibitem{sarnek}
A. M. Wusatowska-Sarnek, H. Miura, and T. Sakai, 
J. Mater. Sci. {\bf 34}, 5477 (1999).

\bibitem{note1}
Although edge dislocations can also avoid impenetrable precipitates by the local climb motion, 
this process requires long-range self-diffusion and is irrelevant to materials at room temperatures.

\bibitem{finnis}
M. W. Finnis and J. E. Sinclair, Phil. Mag. A {\bf 50}, 45 (1984).

\bibitem{ackland}
G. J. Ackland, D. J. Bacon, A. F. Calder, and T. Harry, 
Phil. Mag. A {\bf 75}, 713 (1997).

\bibitem{li}
J. Li, Modelling Simul. Mater. Sci. Eng. {\bf 11}, 173 (2003).

\bibitem{hatano}
T. Hatano and H. Matsui, Phys. Rev. B {\bf 72}, 094105 (2005).

\bibitem{sato}
Y. Satoh, T. Yoshiie, H. Mori, and M. Kiritani, 
Mater. Sci. Eng. {\bf A 350}, 44 (2003).

\bibitem{brown3}
L. M. Brown, Phil. Trans. R. Soc. Lond. A {\bf 355}, 1979 (1997).

\bibitem{asymmetry}
Although the model precipitate of immobile atoms is somewhat artificial, the asymmetric shear stress
may be realized if other lattice defects exist near the precipitate, such as dislocations or grain boundary. 

\bibitem{zbib}
H. M. Zbib, M. Rhee, J. P. Hirth, and T. D. de La Rubia,
J. Mech. Behavior Mater. {\bf 11}, 251 (2000).

\bibitem{kardar}
M. Kardar, in {\it Fundamental Problems in Statistical Mechanics}, edited by H. van Beijeren,
Phys. Rep. {\bf 301}, 85 (1998).

\end{thebibliography}
\end{document}